\documentclass[aps,prd,twocolumn,groupedaddress,nofootinbib]{revtex4-2}  

\usepackage{graphicx}  
\usepackage{dcolumn}   
\usepackage{bm}        
\usepackage{amssymb,amsfonts,amsmath,physics}   
\usepackage{cancel}
\usepackage{mathtools}
\usepackage{slashed}

\usepackage{tikz-feynman}
\tikzfeynmanset{compat=1.1.0}
\usetikzlibrary{arrows}
\tikzset{	
	vertex/.style={circle,draw, minimum size=1.5em},	
	edge/.style={->,> = latex'}	
}
\usepackage[bookmarks, breaklinks, colorlinks,urlcolor=blue, citecolor=red, 
linkcolor=blue]{hyperref}
\usepackage[normalem]{ulem}

\newcommand{\be}{\begin{eqnarray*}}
	\newcommand{\ee}{\end{eqnarray*}}

\newcommand{\bee}{\begin{eqnarray}}
	\newcommand{\eee}{\end{eqnarray}}
\newcommand{\beeq}{\begin{equation}}
	\newcommand{\eeq}{\end{equation}}

\usepackage{xspace}

\newcommand{\ba}{\begin{array}}
	\newcommand{\ea}{\end{array}}
\newcommand{\bd}{\begin{displaymath}}
	\newcommand{\ed}{\end{displaymath}}
\newcommand{\besub}{\begin{subequations}}
	\newcommand{\eesub}{\end{subequations}}
\newcommand{\bea}{\begin{eqnarray}}
	\newcommand{\eea}{\end{eqnarray}}

\def\q2 {q^2}

\usepackage{tikz}
\usetikzlibrary{arrows,shapes}
\usetikzlibrary{trees}
\usetikzlibrary{matrix,arrows} 				
\usetikzlibrary{positioning}				
\usetikzlibrary{calc,through}				
\usetikzlibrary{decorations.pathreplacing}  
\usepackage{pgffor}							

\usetikzlibrary{decorations.pathmorphing}	
\usetikzlibrary{decorations.markings}
\tikzset{
	vector/.style={decorate, decoration={snake}, draw},
	provector/.style={decorate, decoration={snake,amplitude=2.5pt}, draw},
	antivector/.style={decorate, decoration={snake,amplitude=-2.5pt}, draw},
	fermion/.style={draw=black, postaction={decorate},
		decoration={markings,mark=at position .55 with {\arrow[draw=black]{>}}}},
	fermionbar/.style={draw=black, postaction={decorate},
		decoration={markings,mark=at position .55 with {\arrow[draw=black]{<}}}},
	fermionnoarrow/.style={draw=black},
	gluon/.style={decorate, draw=black,
		decoration={coil,amplitude=4pt, segment length=5pt}},
	scalar/.style={dashed,draw=black, postaction={decorate},
		decoration={markings,mark=at position .55 with {\arrow[draw=black]{>}}}},
	scalarbar/.style={dashed,draw=black, postaction={decorate},
		decoration={markings,mark=at position .55 with {\arrow[draw=black]{<}}}},
	scalarnoarrow/.style={dashed,draw=black},
	electron/.style={draw=black, postaction={decorate},
		decoration={markings,mark=at position .55 with {\arrow[draw=black]{>}}}},
	bigvector/.style={decorate, decoration={snake,amplitude=4pt}, draw},
}

\tikzstyle{block} = [draw, rectangle, 
minimum height=3em, minimum width=6em]

\begin{document}
\title{Reconciling ALP Dark Matter and Electroweak Baryogenesis through First-Order Electroweak Phase Transition}

\author{Dipendu Bhandari}
\email{dbhandari@iitg.ac.in}
\affiliation{Department of Physics, Indian Institute of Technology Guwahati, Assam-781039, India}

\author{Soumen Kumar Manna}
\email{skmanna2021@gmail.com}
\affiliation{Department of Physics, Indian Institute of Technology Bombay, Mumbai 400076, India}

\author{Arunansu Sil}
\email{asil@iitg.ac.in}
\affiliation{Department of Physics, Indian Institute of Technology Guwahati, Assam-781039, India}

\begin{abstract} 

We show that an axionlike particle (ALP) can simultaneously generate the baryon asymmetry and constitute dark matter through dynamics triggered by a first-order electroweak phase transition (EWPT). In our proposal, the transition briefly reshapes the ALP potential via a temperature-dependent vacuum expectation value of a scalar field $S$, responsible for making the EWPT of first order, inducing a transient mass enhancement of ALP via higher-dimensional $U(1)$-breaking operator(s). This sudden kick generates a large ALP velocity near the onset of EWPT enabling the broadening of relic satisfied parameter space and predict a complementary stochastic gravitational-wave signal from the underlying first-order transition. We further show that the same ALP dynamics can naturally fuel electroweak baryogenesis through its coupling to electroweak anomaly.

\end{abstract}
\maketitle
\section{Introduction}

The existence of baryon asymmetry of the Universe (BAU)~\cite{Planck:2018vyg} remains one of the long-standing problems within the Standard Model (SM). An underlying key to address this issue is perhaps to connect its genesis with the period of electroweak phase transition (EWPT) termed as electroweak baryogenesis (EWBG), particularly because the {\it{baryon-number violation}} in SM (a crucial ingredient to realise BAU) via electroweak sphaleron process gets decoupled during this epoch only. However, within the SM framework, the EWPT is a smooth {\it cross-over}~\cite{DOnofrio:2015gop} rather than a strong first-order phase transition (FOPT), precluding the other requirement for obtaining BAU: {\it{the departure from thermal equilibrium}}. Moreover, the {\it{CP violation}} within the SM falls far below the requirement. These shortcomings necessitate the introduction of new physics beyond the SM (BSM) to provide a strong FOPT and simultaneously introduce new sources of CP violation to realize EWBG.

Alongside, dark matter (DM)~\cite{Julian:1967zz,SDSS:2003eyi} of the Universe poses another central open question in particle physics and cosmology, a resolution of which requires physics beyond the SM too. While weakly interacting massive particles (WIMPs)~\cite{Bertone:2004pz,Feng:2010gw,Bergstrom:2012fi} were long considered leading candidates, increasingly stringent null results from direct-detection experiments \cite{LZ:2024zvo} have shifted attention toward lighter, more feebly interacting states. Among these, the QCD axion and axion-like particles (ALPs) are well-motivated dark matter candidates, particularly through the misalignment mechanism~\cite{Preskill:1982cy,Abbott:1982af,Dine:1982ah,Turner:1983he,Arias:2012az}. The axion field $a$, initially displaced from its minimum during inflation, begins coherent oscillations (with a nonzero velocity $\dot{a}$) once the Hubble expansion rate ($\mathcal{H}$) falls below the axion mass ($m_a$), after which its energy density redshifts as non-relativistic matter, $\rho_a \propto R^{-3}$.

In this work, we investigate whether an axion or ALP can play a dynamical role in electroweak baryogenesis while simultaneously accounting for the observed dark matter relic abundance. This question is motivated by the observation that the onset of coherent axion/ALP oscillations (resembling as DM) generically features a nonzero field velocity, which in turn can induce a CPT-violating effective chemical potential. Such a chemical potential ($\mu_{\rm eff}$) arises provided there exists derivative interaction between the axion/ALP and the SM baryon or lepton currents ($j_X^{\mu}$) of the form $\partial_\mu a\, j_X^\mu / f_a$, which reduces to $\dot{a}\, j_X^0 / f_a \equiv \mu_{\rm eff}$, after axion/ALP gains a nonzero velocity, under the assumption that the axion/ALP field is spatially homogeneous. This required homogeneity is naturally realised if the global $U(1)$ symmetry associated with the axion/ALP is spontaneously broken before inflation, while the second necessary condition, an efficient baryon-number-violating process in equilibrium, is provided by electroweak sphalerons itself prior to their freeze-out. Together, these ingredients hint toward a plausible avenue for axion or ALP-assisted baryogenesis, through electroweak (EW) anomaly, compatible with dark matter phenomenology.

It can therefore be readily understood that the standard QCD axion is not the right candidate in this regard, as the PQ axion starts gaining nonzero velocity around a temperature of the Universe, close to the QCD condensate (far below the EWPT) where the electroweak sphaleron (source of baryon number violation) already decoupled. 
Nevertheless, in certain nonstandard post-inflationary scenarios, such as axiogenesis~\cite{Co:2019wyp} or setups involving kinetic misalignment~\cite{Co:2020jtv,Datta:2024xhg}, the axion (or a ALP) can attain an early nonzero velocity, allowing it to participate in baryogenesis. Conversely, if the QCD axion is required to account for the observed dark matter relic abundance, one typically needs a decay constant $f_a \sim 10^{12}\,\mathrm{GeV}$ for an initial misalignment angle $\theta_i = a_i / f_a \sim \mathcal{O}(1)$ \cite{DiLuzio:2020wdo} which corresponds to the onset of axion oscillation around $T\sim$ 1 GeV. This follows from the intrinsic relation between the axion mass $m_a$ and $f_a$, set by the QCD scale $\Lambda_{\rm QCD} \sim 150\,\mathrm{MeV}$~\cite{DiLuzio:2020wdo}. Taken together, these considerations make it difficult for the minimal QCD axion framework to simultaneously realize both electroweak baryogenesis and dark-matter phenomenology in a realistic way.

A recent work~\cite{Bhandari:2025pzg} of us has demonstrated that a first-order EWPT (FOEWPT) can significantly influence the dynamics of the QCD axion. This effect called {\it recurrent misalignment} arises from a higher-dimensional, $U(1)_{\rm PQ}$-breaking interaction between the global symmetry breaking scalar $\Phi$ and a real SM singlet scalar $S$, enabling the EWPT of first order via its SM Higgs ($H$) portal interaction, introduced through a portal operator $S^{4}\Phi^{2} e^{i\alpha}/{\Lambda^{2}} + \mathrm{h.c.}$ having $\Lambda$ as the cut-off scale. The scalar $S$ acquires a vacuum expectation value ({\it vev}) at a temperature $T_s$, generating a tree-level potential barrier between the symmetric and the broken electroweak phase. The brief involvement of $\langle S\rangle$ across the transition imparts a transient axion mass prior to the onset of QCD confinement, allowing axion oscillations to begin during the electroweak epoch. This mechanism is shown to substantially modify the conventional QCD axion phenomenology, particularly its dark matter relic abundance, with the scale $\Lambda$ playing a crucial role.

A natural question then is whether such a FOEWPT framework can simultaneously account for both the observed baryon asymmetry and the dark matter abundance through axion dynamics. For the QCD axion, the tight relation between its mass and decay constant, further restricted by astrophysical constraints such as $f_a \gtrsim 10^8$ GeV from SN1987A~\cite{Raffelt:2006cw}, leaves no remaining room to accommodate the required baryon asymmetry while also satisfying the relic density constraint; achieving the latter inevitably suppresses the former. In contrast, ALPs evade this restriction because their mass and decay constant are independent parameters. We demonstrate here that the same non-renormalizable operator, responsible for recurrent misalignment by inducing a transient, enhanced ALP mass during the FOEWPT, can in turn result into the large ALP velocity exactly when EW sphalerons are active, allowing efficient baryon asymmetry generation. Subsequently, as the Universe cools below the electroweak scale, the ALP mass relaxes to its near-original value (barring an additional contribution from non-renormalizable SM Higgs portal interaction), and its oscillations provide the correct dark matter relic density through the {\it{recurrent misalignment}} mechanism. Thus, the transient dynamics of the ALP field during the EWPT provides a unified origin for both dark matter and the baryon asymmetry of the Universe.

There are several studies in the literature exploring ALPs in connection with the EWPT. In~\cite{Jeong:2018jqe,Harigaya:2023bmp} and~\cite{Jeong:2018ucz}, a first-order EWPT is triggered by the ALP–Higgs coupling, while the baryon asymmetry is generated through the effective coupling of the ALP to the EW anomaly and to the top quark Yukawa interaction, respectively; however, these works do not address ALP dark matter. In~\cite{Im:2021xoy}, the ALP explains the baryon asymmetry via its effective coupling to the EW anomaly within a clockwork framework, and the same ALP constitutes the dark matter through its interaction with a hidden gauge sector, although in this case the electroweak phase transition remains a crossover. In another study~\cite{Jeong:2024hhi}, the baryon asymmetry is generated through the ALP–EW anomaly coupling during a first order EWPT induced by an ALP–Higgs interaction, while the dark matter is accounted for by axion-coupled dark photons. In contrast, our scenario brings together ALP dark matter, electroweak baryogenesis, and the first-order EWPT, while enlarging the viable ALP dark matter parameter space in light of existing constraints and providing complementary connections to the stochastic gravitational wave (GW) signal from the underlying FOPT.

\section{The Framework}
In this work, we extend our earlier framework of QCD axion to an ALP case which would not only provide the correct DM relic abundance but also connecting it with a spontaneous electroweak baryogenesis~\cite{Cohen:1987vi,Cohen:1988kt,Cohen:1990py,Cohen:1991iu,Nelson:1991ab,DeSimone:2016ofp} scenario across FOEWPT. In doing so, we first specify in this section its relevant effective interactions to SM and other BSM fields so as to identify the epoch of its oscillation across the EWPT, with a particular emphasis on being of first order. Thereafter, we proceed to discuss the mechanism for generating the DM relic abundance and BAU using ALP dynamics (using these interactions) and associated phenomenological discussion in subsequent sections. 

We presume that ALPs exist prior to the end of primordial inflation due to the spontaneous breaking of a global $U(1)$ symmetry during inflation. After inflation and reheating, the ALP field becomes spatially homogeneous, $a(t)$, and remains frozen at an initial value $a_I$, set by the misalignment angle $\theta_i = a_i/f_a$ with $\dot{\theta}_i = 0$, as long as the Hubble $\mathcal{H} \gg m_{a0}$. As a Nambu-Goldstone boson, the ALP acquires its mass $m_{a0}$ from shift-symmetry breaking contributions, likely through a hidden sector non-perturbative dynamics analogous to the QCD axion. 

Since we are interested to investigate the impact of EWPT being first order on such ALPs, we introduce a SM real singlet scalar field $S$, odd under a $Z_2$ symmetry, which can trigger \cite{Espinosa:2011ax} a first-order electroweak phase transition through its SM Higgs-portal interaction 
\begin{align}
	-\mathcal{L}_{\rm{PT}} = \lambda_{hs} |H|^2 S^2.
	\label{eq:Higgs-portal}
\end{align}
For a suitable choice of $\lambda_{hs}$ (as we show in a later section), the {\it vev} of $S$ at a critical temperature $T_c$ can generate a barrier between the phases [$\langle H \rangle = 0; \langle S \rangle = v_s$] and [$\langle H \rangle = v_h; \langle S \rangle = 0$]. In order such FOEWPT to be impacted upon the dynamics of ALP, we consider the presence of higher-dimensional operators\footnote{A possible origin for these operators is discussed in the appendix of Refs.~\cite{Bhandari:2025pzg,Manna:2023zuq}.}~\cite{Bhandari:2025pzg,Manna:2023zuq}
\begin{align}
	-\mathcal{L}^{\rm ALP}_{\rm{int}} = \frac{S^4}{\Lambda^2} \Phi^2 e^{i\alpha} + \kappa \frac{|H|^4}{\Lambda^2} \Phi^2 e^{i\beta} + h.c. ,
	\label{eq:dim-6-explicit}
\end{align}
where the $\Phi~(=\eta e^{ia/\sqrt{2}f_a})$ field spontaneously breaks the global $U(1)$ symmetry, resulting ALP $(a)$ as the pNGB and $\Lambda$ denotes the cut-off scale of the theory. The radial field $\eta$ is assumed to be very heavy so as not to contribute in the physics below the scale $\Lambda$. One can expect such explicit global $U(1)$ breaking interactions, suppressed by the Planck scale $(M_{\rm Pl})$ from quantum gravity effects \cite{Giddings:1988cx,Coleman:1988tj,Rey:1989mg,Abbott:1989jw,Akhmedov:1992hh,Kamionkowski:1992mf,Kallosh:1995hi}. However, as shown in Ref. \cite{Draper:2022pvk,Cordova:2022rer}, the cut-off scale $(\Lambda)$ can be well-below the Planck scale $(\Lambda\lesssim M_{\rm Pl})$ following the weak-gravity conjecture \cite{Draper:2022pvk,Cordova:2022rer}. Here, $\kappa$ refers to a dimensionless coupling that measures the relative strength between the two explicit breaking terms. Also, $\alpha$ and $\beta$ in Eq. \ref{eq:dim-6-explicit} denote the additional phases, which we will fix at the value of $\pi$ throughout this work, simplifying our analysis without any alteration in the ALP potential minimum\footnote{The impact of choosing phases different from 
$\pi$ is explored in detail in Ref.~\cite{Manna:2023zuq}.}. The detailed effect of such interactions in altering the ALP mass (during EWPT) and its dynamics to predict the DM relic abundance will be studied in detail in the following sections. 

In this work, as we plan to extend the {\it{recurrent misalignment}} mechanism \cite{Bhandari:2025pzg} to further connect it with a spontaneous electroweak baryogenesis, a key consideration would be to introduce an effective interaction between ALP and the electroweak anomaly:
\beeq
-\mathcal{L}^{\rm ALP}_{\rm{eff}} =  \frac{1}{16 \pi^2} \frac{a}{f_a} {\rm Tr} W^{\mu\nu}\tilde{W}_{\mu\nu},
\label{EW anomaly-ALP}
\eeq
which generates a chemical potential after ALP starts oscillating. This in turn induces a net baryon asymmetry in presence of sphaleron process being in equilibrium, during FOEWPT, which requires a sizable chemical potential, or equivalently, a large axion velocity. The exact details are included in a dedicated section on baryogenesis. Note that such an effective interaction is reminiscent of loops of heavy leptons charged under the same $U(1)$. Although such a generation of BAU sounds similar to spontaneous baryogenesis via leptogenesis~\cite{Li:2001st,Yamaguchi:2002vw,Kusenko:2014uta,Ibe:2015nfa,Takahashi:2015ula,Bae:2018mlv,Domcke:2020kcp,Berbig:2023uzs,Chao:2023ojl,Chun:2023eqc,Chun:2025abp,Chun:2024gvp} scenario, there are salient features associated with our proposal. For example, in \cite{Kusenko:2014uta,Ibe:2015nfa,Bae:2018mlv}, lepton number violation follows from a dimension-five Weinberg operator, responsible for neutrino mass generation\footnote{Contrarily, in~\cite{Datta:2024xhg}, some of us have pointed out that introduction of a different Weinberg-like operator can make it interesting by lowering the scale of asymmetry generation down to electroweak scale.}, needs to be in thermal equilibrium and as a result, the asymmetry generation can only be operative at scales ($T \sim 10^{13}$ GeV) much above the electroweak epoch. In contrast, here we explore ALP-induced baryogenesis mechanism during the EWPT, $i.e.$ at a much low scale making it interesting from experimental point of view as well. Furthermore, the same ALP being responsible for DM, an interesting correlation will follow as we explore in the following sections along with the GW signal related to FOEWPT.

\section{First Order Electroweak phase transition with real singlet}

Before going into the details of ALP dynamics as impacted by the first order EWPT, in this section, we provide a brief overview of first order EWPT and related parameters. This will be helpful in governing the thermal evolution of the $vev$s 
of the real singlet $S$ and Higgs fields ($v_s$, $v_h$) while simultaneously fixing the nucleation temperature $T_n$. To begin, we write down the $Z_2$-symmetric tree level potential involving the SM Higgs and the singlet $S$ field (inclusive of the Lagrangian of Eq. \ref{eq:Higgs-portal}) as~\cite{Espinosa:2011ax}:
\begin{align}
		V_0(H,S) =\, & \mu_h^2(H^{\dagger} H) + \lambda_h(H^{\dagger} H)^2 \nonumber \\ 
		&  + \frac{\mu_s^2}{2} S^2 + \frac{\lambda_s}{4} S^4 + \frac{\lambda_{hs}}{2} (H^{\dagger} H) S^2 , \label{tree-level potential}
\end{align}
where, $\mu_h^2, \mu_s^2 <0.$ One also needs to incorporate the thermal corrections to construct the effective potential $V_{\rm eff}$. In the high-temperature approximation, retaining the leading $T^2$-thermal corrections and neglecting subleading one-loop Coleman–Weinberg terms, one obtains (see Appendix~\ref{ap:effective-potential})
\begin{align}
	V_{\rm eff}(h,S,T) =\, &  \frac{1}{2}( \mu_h^2 + c_h T^2) h^2 + \frac{\lambda_h}{4} h^4 + \nonumber \\
	& \frac{1}{2}( \mu_s^2 + c_s T^2) S^2 + \frac{\lambda_s}{4} S^4 +  \frac{\lambda_{hs}}{4} h^2 S^2,     \label{effective potential}
\end{align}
where $h$ denotes the classical background of the SM Higgs field and $c_h$ and $c_s$ are defined as:
\begin{align}
	c_h =\, &  \frac{1}{48}(9 g_{\rm w}^2 +3 g_Y^2 +12 y_t^2 + 24 \lambda_h + 2 \lambda_{hs}),  \\
c_s =\, &	\frac{1}{12}(3 \lambda_s + 2 \lambda_{hs}).
\end{align}

In this setup, the EWPT proceeds in two steps. At high temperatures, all symmetries are restored with $\langle H\rangle=\langle S\rangle=0$. Then, as the temperature goes down to $T_s$, the $S$ field starts to develop a temperature dependent {\it vev},
\begin{align}
	\langle S \rangle \equiv v_{s}(T) = \sqrt{-\frac{\mu_s^2 + c_s T^2}{\lambda_s}},
	\label{s-vev}
\end{align}
while $\langle H \rangle=0$. This stage corresponds to spontaneous breaking of $Z_2$ symmetry and modifies the ALP mass via higher-dimensional operators, to be explicitly shown in the next section. As the temperature decreases, the Higgs vacuum develops a new minimum with
\begin{align}
	\langle h \rangle \equiv v_h(T) = \sqrt{-\frac{\mu_h^2 + c_h T^2}{\lambda_h}},
	\label{h-vev}
\end{align}
while $vev$ of the $S$ field vanishes, $\langle S\rangle=0$. In between, there exists a critical temperature $T_c ~(< T_s)$ when the Higgs minimum becomes degenerate with the singlet minimum, with the appearance of potential barrier between them, which marks the onset of a first-order phase transition.

For $T< T_c$, the electroweak minimum with $v_h(T)\neq 0$ and $v_s(T)=0$ becomes the true vacuum. The transition proceeds through bubble nucleation at the nucleation temperature $T_n$, governed by
\begin{align}
	\Gamma_d(T_n) \simeq \mathcal{H}(T_n)^4 ,
	\label{Nucleation condition}
\end{align}
with $\mathcal{H}(T)=1.66\sqrt{g_\star(T)}T^2/M_{\rm Pl}$, where the vacuum decay rate is~\cite{Coleman:1977py,Linde:1980tt,Linde:1981zj} 
\begin{align}
	\Gamma_d(T) \simeq T^4 \left( \frac{S_3}{2\pi T} \right)^{3/2} e^{-S_3/T}.
\end{align}
Here, the 3-dimensional Euclidean action for $O(3)$-symmetric bounce~\cite{Linde:1981zj,Coleman:1977py} is denoted by $S_3$ (which we calculated by the {\bf FindBounce}~\cite{Guada:2020xnz} package, as discussed in Appendix~\ref{ap:Action-calculation}).
 \begin{table}[!htb]
	\begin{tabular}{| c | c | c | c || c| c | c |}
		\hline
		& $\lambda_{s}$  & $\lambda_{hs}$ & $\mu_s^2$ (GeV$^2$) & $T_s$ (GeV) & $T_c$ (GeV) & $T_n$ (GeV)  \\
		\hline
		BP1 & 0.03 &  0.22  &  -2240.5  &  225.23  &  100  &  42.7   \\
		\hline
		BP2 & 0.1 &  0.37  &  -4367.5  &  224.49  &  95  &  51.4   \\
		\hline
	\end{tabular}
	\caption{Benchmark points satisfying FOEWPT}
	\label{tab:Two Benchmark}
\end{table}

In Table~\ref{tab:Two Benchmark}, we list two benchmark choices for the parameters $\lambda_s, \lambda_{hs}$, and $\mu_s^2$ which determine the values $T_s, T_c$ and $T_n$, ensuring both a FOEWPT and its successful completion. For the rest of our analysis, we adopt these benchmark points.
We also note that the singlet scalar field $S$, being odd under $Z_2$, becomes stable, but its contribution to the dark matter relic density is negligible, largely because its Higgs-portal interaction is relatively strong.

\section{Evolution of ALP as Dark Matter}

In this section, we examine the evolution of the ALP mass in presence of the dimension-6 operators of Eq.~\ref{eq:dim-6-explicit} and assess their implications for ALP dynamics. The ALP acquires a bare mass $m_{a0}$ possibly from a non-perturbative potential of the form $ V_0 (a) = m_{a0}^2 f_a^2 \left[1 - \cos(a/f_a)\right]$, generated by a QCD-like hidden sector. At early times, the field remains frozen at an initial misalignment angle $\theta_i \sim 1$, with Hubble friction preventing coherent motion until the expansion rate drops below the mass scale. Only then does the ALP commence oscillations about its potential minimum. The dimension-6 operators can modify the ALP mass only in the vicinity of the electroweak phase transition (EWPT), and understanding their impact requires determining whether oscillations begin (the indicative temperature of which is $T_{\rm osc}$) with the bare mass $m_{a0}$ at temperatures above or below $T_s$. If oscillations do start early, the subsequent mass enhancement across the FOEWPT would alter the evolution of the field amplitude and energy density. Conversely, if the ALP remains over-damped until temperatures below the EWPT, the onset of oscillations may instead coincide with $T_s$, leading to a qualitatively different dynamical history. 

\begin{figure}[htb!]
	\includegraphics[width=\columnwidth]{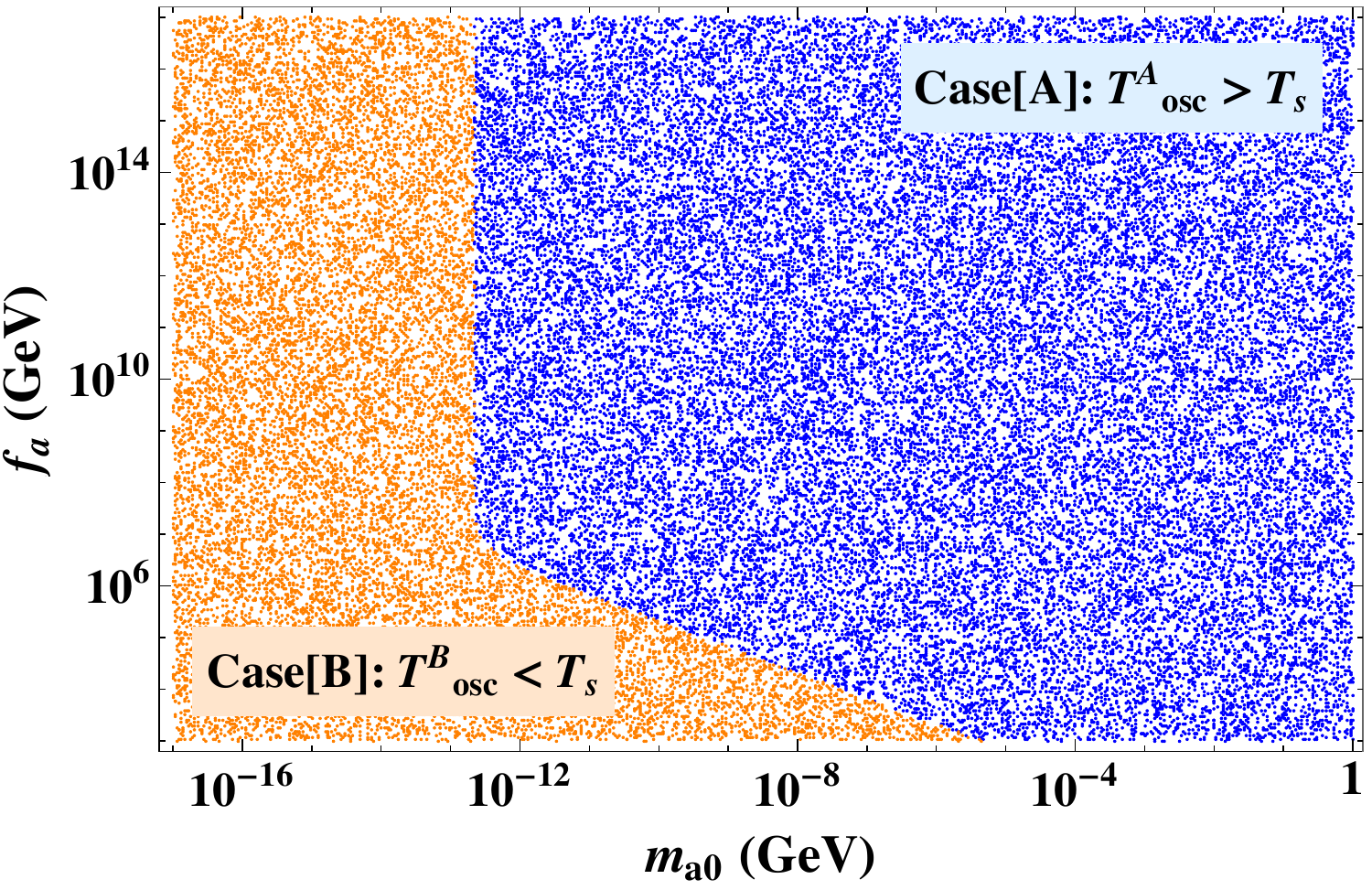}
	\caption{ Regions in the $m_{a0}-f_a$ plane showing whether ALP oscillations begin above (blue) or below (orange) the temperature $T_s$.}
	\label{fig:ALP-ma-fa-oscillationTemp}
\end{figure}

Hence, before entering into the details of ALP mass-variation, let us first analyse whether the ALP starts its oscillation with its bare mass $m_{a0}$ ($i.e.$, without ALP mass variation across the FOEWPT), prior to the temperature $T_s$ or not. In this context, we note that an additional thermal friction ($\Upsilon_{a}$), arising from the interaction of ALP with the electroweak anomaly as in Eq.~\ref{EW anomaly-ALP}, acts on the ALP in this set-up. Taking this into consideration, we specify two cases, namely
\begin{eqnarray}
	{\rm Case \text{-}A:}& T_{\rm osc}^A > T_s,  \nonumber \\
	{\rm Case\text{-}B:}& T_{\rm osc}^B \leq T_s.
	\nonumber
\end{eqnarray} 
Note that the demarcation between these two cases is set by the condition,
\begin{equation}
	m_{a0} = 3 \mathcal{H}(T) + \Upsilon_{a}(T) , \label{eq:oscillation-cases-condition1}
\end{equation}
where, the thermal friction term $ \Upsilon_{a}(T)$, induced through the coupling of ALP with the SM gauge sector, is given by~\cite{Im:2021xoy}
\begin{equation}
	 \Upsilon_{a}(T) =  \frac{\Gamma_{\rm sp}}{2 T f_a^2}, \label{eq:Thermal friction}
\end{equation}
with $\Gamma_{\rm sp}\simeq 18 \alpha_{\rm W}^5 T^4$~\cite{DOnofrio:2014rug} be the sphaleron rate in the symmetric phase of the Universe.  We see from Eq.~\ref{eq:oscillation-cases-condition1} that the oscillation temperature depends on both $m_{a0}$ and $f_a$, hence we cannot divide the two regimes of whether the oscillation temperature below or above $T_s$ solely on the basis of $m_{a0}$ but there is $f_a$ dependence also. We have shown in Fig.~\ref{fig:ALP-ma-fa-oscillationTemp} the two regions in $m_{a0}-f_a$ plane signifying the oscillation temperature above or below $T_s$. 
We observe in Fig.~\ref{fig:ALP-ma-fa-oscillationTemp} that the vertical dividing line between the blue and orange regions bends toward larger values of $m_{a0}$ when $f_a$ falls below a certain critical value. The reason is that for $f_a$ below this threshold, the thermal friction $ \Upsilon_{a}(T)$ overtakes the Hubble friction. Consequently, even for relatively large $m_{a0}$, the ALP cannot begin oscillating above $T_s$. This leads to an extended region where the oscillation temperature lies below $T_s$ in contrast to the standard expectation in which larger $m_{a0}$ would typically initiate oscillations earlier.
The blue and orange regions with [$m_{a0} - f_a$] correlation, corresponds to case-A and case-B respectively, have been used to further study the ALP as dark matter taking into account the mass-enhancement across FOEWPT, in the subsequent discussion.

We now turn our attention to the ALP mass variation, beyond its bare mass $m_{a0}$, in view of Eq. \ref{eq:dim-6-explicit}. To streamline the discussion and highlight the relevant physical regimes, we divide the analysis into three 
phases:\\

\noindent [Phase-{\bf I}] Above $T_s$, both $H$ and $S$ have vanishing {\it vevs}. Therefore, in this regime, ALP only has its original mass $m_{a0}$. \\
\noindent [Phase-{\bf II}] Between $T_s$ and $T_n$, the nucleation temperature of EWPT : In this period, the $S$ field starts to develop non-zero {\it vev}, $\langle S\rangle=v_s(T)$, yielding an additional contribution to the ALP mass. \\
\noindent [Phase-{\bf III}] Below $T_n$: The {\it vev} of the $S$ field vanishes, while the SM Higgs vacuum settles into a non-zero value $\langle H\rangle=v_h(T)$, which approaches $v_h=246$ GeV at zero temperature. Thus, below $T_n$, the final ALP mass receives contributions only from the Higgs-portal term, not from the $S$-portal. The variations of ALP mass can be summarized as
\begin{widetext}
	\begin{equation}
		m_a (T) =
		\begin{cases}
			m_{a0} & \text{;} \quad T>T_s, \\
			\left[ m_{a0}^2  + \frac{4 v_s^4(T)}{\Lambda^2}\right]^{1/2} & \text{;}  \quad  T_n < T \leq T_s,\\
			\left[ m_{a0}^2 + \kappa \frac{v_h^4(T)}{\Lambda^2}\right]^{1/2} & \text{;}  \quad T\leq T_{n}.
		\end{cases}
		\label{eq:mass-variation}
	\end{equation}
\end{widetext}

We now examine ALP dynamics with the temperature-dependent mass variation across the FOEWPT, accounting the initial conditions similar to the standard misalignment, $\theta_{i} \sim 1$ and hence $\dot{\theta}_{i}=0$. The classical equation of motion of the ALP field $a(T)$ (parametrised by $\theta(T) = a(T)/f_a$) is given by
\begin{equation}
	\ddot{\theta} +( 3 \mathcal{H} + \Upsilon_{a} ) \dot{\theta} + m^2_{a}(T) \sin\theta =0. 
	\label{axion-evolution-eq}
\end{equation}
where ``dot" denotes the derivative with respect to time $t$. Note that, we keep the cut-off scale $\Lambda$ (where $ f_a < \Lambda < M_{\rm Pl}$) as a free parameter in studying the ALP dynamics such that their values can be constrained from DM relic satisfaction. Other parameters involved in the potential $V_{\rm eff}$ are guided by the realisation of FOEWPT. Specifically, we continue the rest of the analysis with BP1 of Table \ref{tab:Two Benchmark}. We proceed below to discuss 
the evolution of the ALP for case A and B, one after other.\\

\subsection{ALP oscillation begins above $T_s$}

Here we discuss Case-A where the ALP starts oscillating prior to $T_s$. Below, we identify the ALP dynamics through three distinct Phases {\bf I}$\rightarrow${\bf III}. \\

\noindent Phase-{\bf I} ($T>T_s$): The ALP mass does not acquire any new contributions to its mass and starts to oscillate at temperature $T_{\rm osc}^A$ upon satisfying the condition 
\beeq
m_{a0} = \left.( 3 \mathcal{H} + \Upsilon_{a} )\right|_{T_{\rm osc}^A}. \label{eq:caseA-Tosc}
\eeq
\vskip -0.25cm
\noindent Phase-{\bf II} ($T_s\geq T > T_n$): In this phase, the $S$ field develops a temperature-dependent $vev ~v_s(T)$. As a result, the axion mass $m_a(T)$ is altered through the portal interaction specified in Eq.~\ref{eq:mass-variation}. By suitably selecting the ratio $v_s(T)/\Lambda$, with $v_s(T)$ determined from the benchmark points in Table~\ref{tab:Two Benchmark}), this additional contribution in ALP mass changes the oscillation frequency in this phase than before. \\
\vskip -0.25cm
\noindent Phase-{\bf III} ($ T \leq T_n$): In this phase, $v_s$ vanishes but the Higgs $vev$, $v_h(T)$ becomes non zero, thereby enabling the Higgs-portal interaction to modify the ALP mass as described in Eq.~\ref{eq:mass-variation}. Again, depending on the choice of $k$, the oscillation frequency of the ALP would change as a result of which we expect a modification in the standard ALP parameter space in $m_a$, $f_a$ plane.

At the oscillation temperature $T_{\rm osc}^A$, the energy density of the ALP is $\rho_a=\frac{1}{2}\left(f_a^2 \dot{\theta}^2 + m_{a0}^2 f_a^2 \theta^2 \right)$. As the ALP does not have any initial velocity ($\dot{\theta_i}=0$), the initial energy density at the onset of oscillation in phase-I is given by
\begin{equation}
	\rho_a (T_{\rm osc}^A) = \frac{1}{2}m_{a0}^2 f_a^2 \theta_i^2  . \label{eq:Initial ALP energy density}
\end{equation}
Below $T_{\rm osc}^A$, the ALP evolve according to the Eq.~\ref{axion-evolution-eq} with decreasing amplitude of oscillation. 
As the ALP comoving number density is conserved~\cite{Arias:2012az}, the ALP energy density at the temperature $T_s$ ($< T_{\rm osc}^A$) is given by
\begin{equation}
	\rho_a(T_s) = \rho_a(T_{\rm osc}^A)\left( \frac{T_s}{T_{\rm osc}^A}\right)^3  \frac{g_{*s}(T_s)}{g_{*s}(T_{\rm osc}^A)}  \frac{m_a(T_s)}{m_{a0}} .   \label{eq:ALP energy density at Ts}
\end{equation}
Below the temperature $T_s$, $v_s(T)$ becomes non zero and gives an additional contribution to the ALP mass. Due to the temperature dependence of $v_s(T)$ (can be seen from Eq.~\ref{s-vev}) , the ALP mass $m_a(T)$ also becomes temperature dependent between $T_s$ to $T_n$. By solving Eq.~\ref{axion-evolution-eq} with such $m_a(T)$, we get the energy density at $T_n$, $\rho_a(T_n)$.

Thereafter, the present day ALP energy density can be estimated as
\begin{equation}
	\rho_a(T_0) = \rho_a(T_n) \left( \frac{T_0}{T_n} \right)^3  \frac{g_{*s}(T_0)}{g_{*s}(T_n)} \frac{m_a(T_0)}{m_a(T_n)},   \label{eq:ALP energy density at Tn}
\end{equation}
where, $T_0 = 2.4 \times 10^{-4}$ eV is the present temperature of the Universe and $g_{*s}(T_0)=3.94$ is the effective number of relativistic $d.o.f$ at the present temperature $T_0$~\cite{Bauer:2017qwy}. The ALP relic density can therefore be determined as
\begin{equation}
	\Omega_a h^2 =\frac{h^2}{\rho_{c,0}} \rho_a(T_0) , \label{eq:Axion-relic-density}
\end{equation}
where, $\rho_{c,0} = 1.05 \times 10^{-5}\,h^2$ GeV ${\rm cm^{-3}}$ denotes the present critical energy density.

\begin{figure}[htb!]
	 \includegraphics[width=\columnwidth]{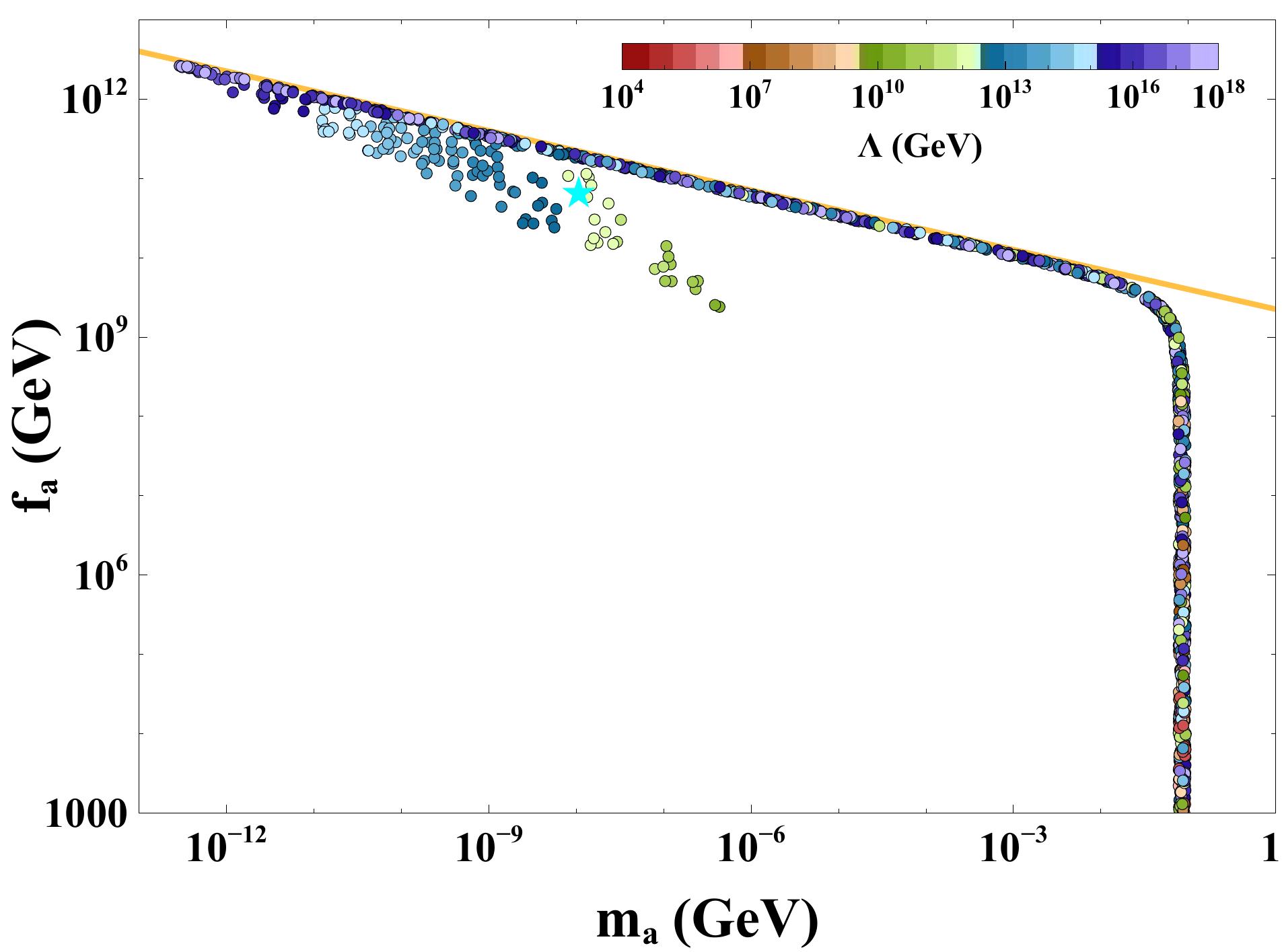}
	\caption{ALP dark matter parameter space in $m_a-f_a$ plane satisfying the correct relic abundance for Case-A. The cyan coloured star indicates the benchmark BP-I in Table~\ref{tab:Two BP-baryonasymmetry-DM relic}, consistent with both the observed BAU and the correct relic density.
		Here, the colorbar indicates the variation of the cut-off scale, $\Lambda$.  }
	\label{fig:ALP-ps-A}
\end{figure}

The effect of such transient ALP mass alteration on its parameter space, where it acts as a dark matter via misalignment mechanism, is illustrated in Fig. \ref{fig:ALP-ps-A} in the $m_a-f_a$ plane with $\Lambda$ (fixed from the relic satisfaction requirement) in the color bar. The blue region of the Fig.~\ref{fig:ALP-ma-fa-oscillationTemp} has been incorporated for scanning of the dark matter relic satisfied parameter space in this case. 
Here (in Fig. \ref{fig:ALP-ps-A}) all the points reproduce the observed dark matter relic abundance $\Omega_a h^2\simeq 0.12$ \cite{Planck:2018vyg}. While the color gradient from dark red to purple indicate the correspondence between $\lbrace m_a, f_a \rbrace$ and the parameter $\Lambda$, with the consideration $f_a\leq \Lambda \leq M_{\rm Pl}$, the narrow orange line along the boundary of the parameter space corresponds to the standard scenario where the ALP mass remains constant and yields the correct relic density. Evidently, in our case, such a contour (orange line) becomes enlarged, which is an artefact of the higher dimensional operators we include in Eq. \ref{eq:dim-6-explicit}. 
Therefore, the relic density not only depends on $m_{a0}$ and $f_a$, but also becomes sensitive to $\Lambda$, providing the additional freedom on the parameter space. 

Assuming that ALP accounts for the entire dark matter abundance, the upper portion of the allowed parameter space (indicated by the band having dotted relic satisfied points) is ruled out due to overproduction of dark matter, while the regions below the allowed band are excluded by the requirement of $\Lambda \gtrsim 2\pi  f_a$. The additional constraint $\Lambda<M_{\rm Pl}$ is relevant mainly for the left edge of the plot, where the smallest possible contribution from the dimension-6 operator to the ALP mass is $v_s(T)^2/M_{\rm Pl}$. We note a sharp fall on the right side of the parameter space in Fig.~\ref{fig:ALP-ps-A}, indicating the insensitivity of the relic density on parameter $f_a$. This is due to the presence of additional thermal friction term $\Upsilon_a$ which imposes $f_a$ dependence on the oscillation temperature $T_{\rm osc}^A$. We can see from Eqs.~\ref{eq:Initial ALP energy density}, \ref{eq:ALP energy density at Ts} and \ref{eq:ALP energy density at Tn} that $\rho_a(T_0)$ proportionals to $m_{a0}, m_a(T_0)$ and the ratio $f_a^2/(T_{\rm osc}^A)^3$.
\begin{figure}[!htb]
	\includegraphics[width=\columnwidth]{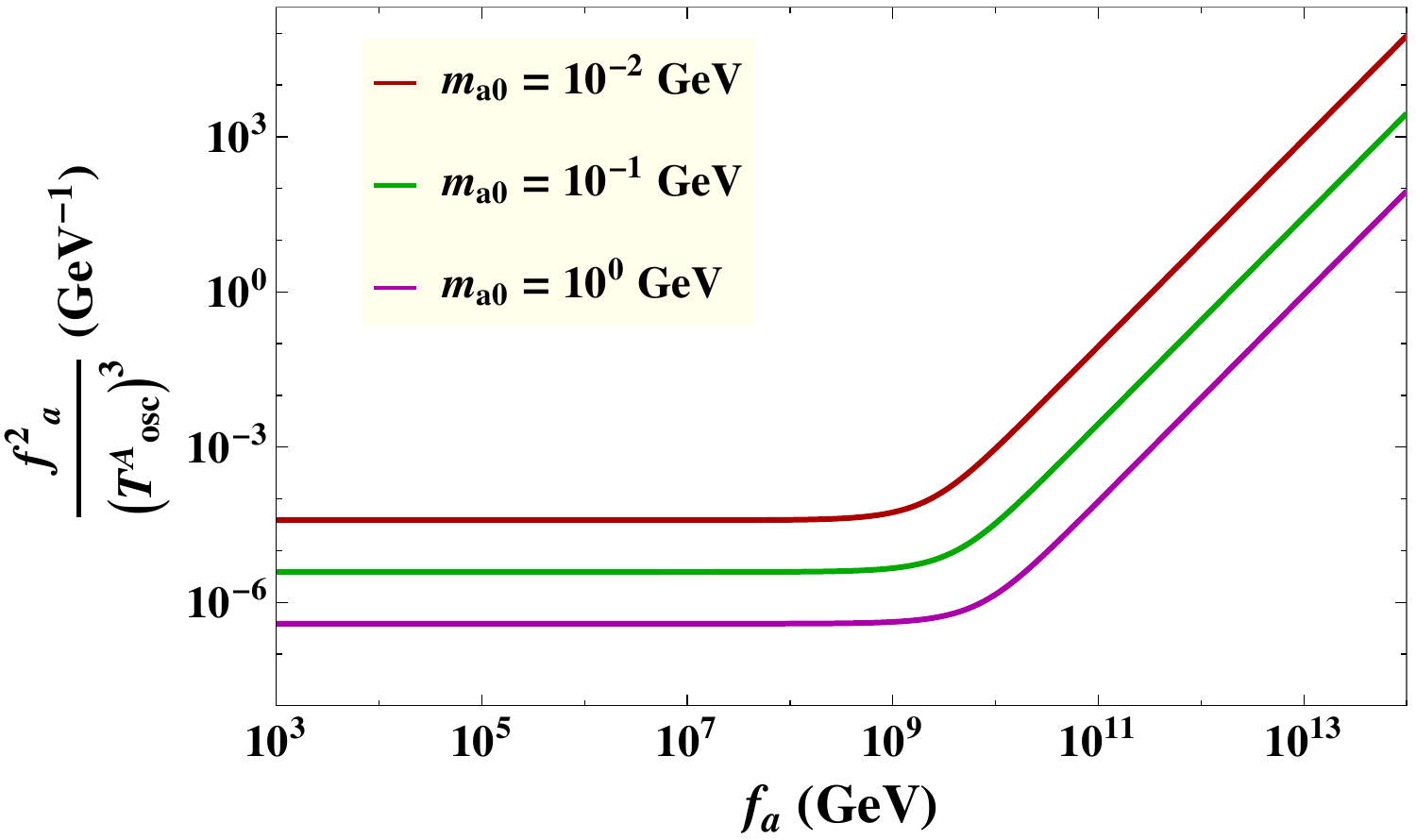}
	\caption{ Variation of $f_a^2/(T_{\rm osc}^A)^3$ with $f_a$ for fixed choices of $m_{a0}$.}
	\label{fig:fa-Tosc}
\end{figure}
 In Fig.~\ref{fig:fa-Tosc}, we have shown that $f_a^2/(T_{\rm osc}^A)^3$ attains a constant value below a certain $f_a$ for three different $m_{a0}$. Again, in the Fig.~\ref{fig:ALP-ps-A}, we see that the drop appears below a certain $f_a$ and makes the relic $f_a$ insensitive as $f_a^2/(T_{\rm osc}^A)^3$ now becomes constant corresponding to a particluar $m_{a0}$. Hence, the final relic can be determined by adjusting the parameter $\Lambda$ while we keep $k=0.1$ constant here. 
 
  \begin{figure}[!htb]
 	\includegraphics[width=\columnwidth]{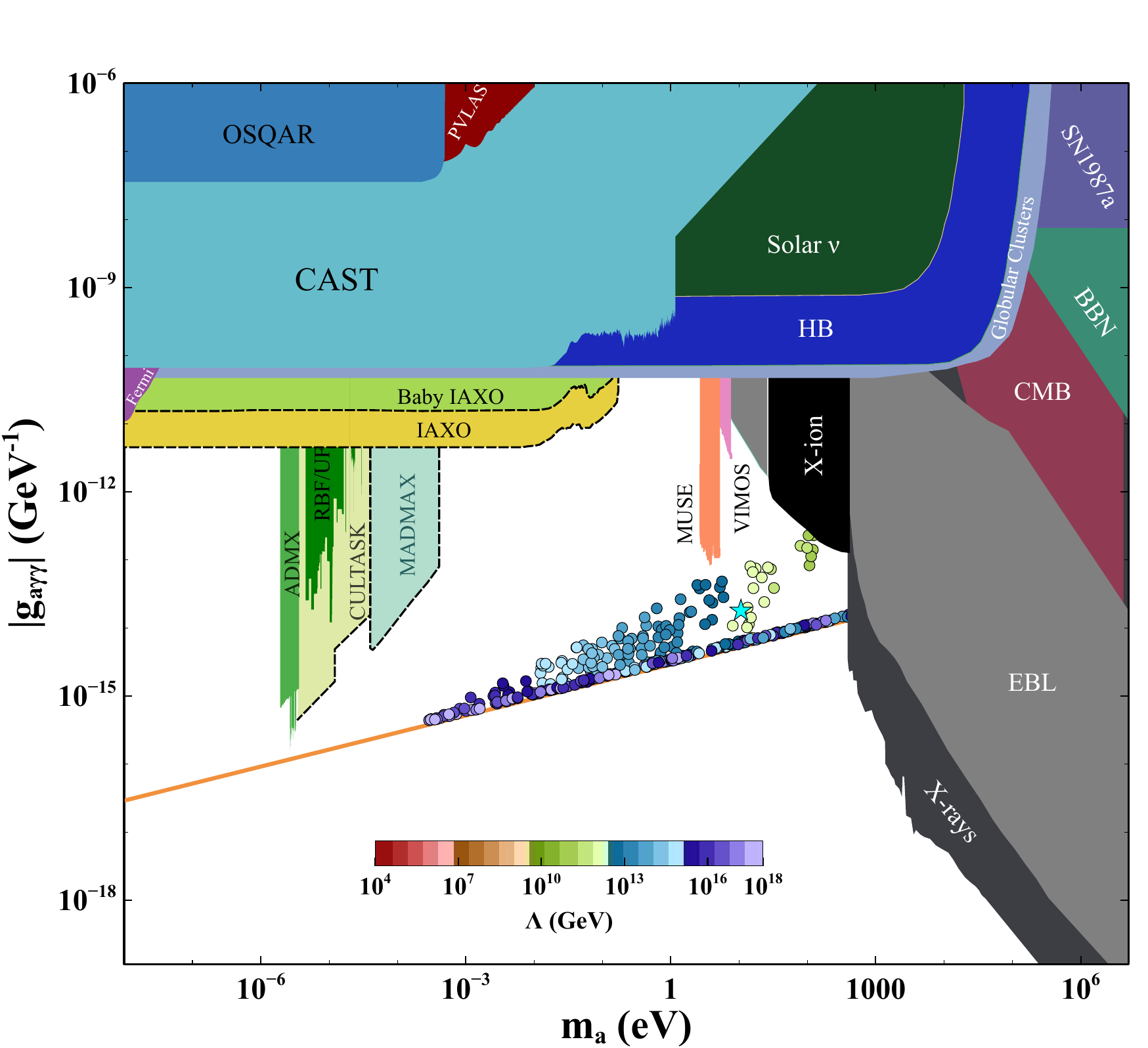}
 	\caption{Excluded regions of the ALP parameter space (in $m_a-g_{a\gamma\gamma}$ plane) from various constraints, together with the relic-density-allowed region for Case-A, adapted from Fig. \ref{fig:ALP-ps-A}. All the experimental, observational and cosmological bounds are taken from the online repository \texttt{AxionLimits} \cite{Ohare2020-gy}.}
 	\label{fig:ALP-photon}
 \end{figure}
 
 Before concluding this subsection, we briefly comment on how this modified ALP parameter space (satisfying correct dark matter relic density) can be relevant from the experimental, observational and cosmological perspective. As shown in Fig.~\ref{fig:ALP-photon}, introducing the effective ALP–photon interaction
 	\beeq
 	\frac{g_{a\gamma\gamma}}{4}aF_{\mu\nu}\tilde{F}^{\mu\nu},
 	\eeq
 	with $g_{a\gamma\gamma}=\frac{\alpha}{2\pi f_a}C_{a\gamma\gamma}$ ($\alpha$ is the fine structure constant and considering $C_{a\gamma\gamma}$ to be $\mathcal{O}$(1)) allows the relic-allowed region to be confronted with the existing and future sensitivities from experiments like haloscopes (such as ADMX \cite{ADMX:2020ote}, MADMAX \cite{Beurthey:2020yuq}), helioscopes (such as CAST \cite{CAST:2017uph}, IAXO \cite{IAXO:2020wwp}) as well as several astrophysical \cite{Raffelt:2006cw,Wouters:2013hua,Ayala:2014pea,Rogers:2020ltq,Reynes:2021bpe,ParticleDataGroup:2022pth} and cosmological sources \cite{Masso:1995tw,Overduin:2004sz,Cadamuro:2011fd,Depta:2020wmr}. Translating the allowed region from the $(m_a,f_a)$ plane into $(m_a,g_{a\gamma\gamma})$ shows that a substantial amount of our extended parameter space falls within the unconstrained white region (see Fig.~\ref{fig:ALP-photon}). Importantly, our parameter space lies below the current CAST and HB limits, while nearly overlapping with the prospective sensitivity of spectroscopic searches using MUSE and VIMOS \cite{Regis:2020fhw}. Therefore, the modified ALP dynamics in this case opens up a significantly broader and experimentally testable ALP parameter space than the standard misalignment scenario.

\subsection{ALP oscillation begins below $T_s$}
Here, the ALP is expected to oscillate at a temperature below the nucleation temprature of EWPT, $T_n$, connected to the mass $m_{a0}$. But in the FOEWPT scenario, due to the presence of those portal interactions as mentioned in Eq.~\ref{eq:dim-6-explicit}, ALP mass is modified in the Phases {\bf I}$\rightarrow${\bf III}, so that its oscillation dynamics can be described as below.\\
\vskip -0.25cm
\noindent Phase-{\bf I} ($T>T_s$): Since the $S$ field $vev$ is zero in this phase, the ALP mass receives no extra contribution from the higher-order PQ-symmetry–breaking term. So the ALP mass is $m_{a0}$ here and by choice of its small value, the ALP cannot oscillate in this phase but instead remains stuck to its initial field value $\theta_i = 1$.\\
\vskip -0.25cm
\noindent Phase-{\bf II} ($T_s\geq T > T_n$): In this phase, the $vev$ of the $S$ field, $v_s(T)$, becomes non zero. Consequently, by the $S$-portal higher-order PQ-symmetry breaking term, the ALP mass gets modified as described in Eq.~\ref{eq:mass-variation}. By making appropriate choice of $v_s(T)/\Lambda$, the ALP starts to oscillate at a temperature $T_{\rm osc}^{B}$ between $T_s$ and $T_n$ directed by $m_{a}(T_{\rm osc}^{B}) \geq \left.( 3 \mathcal{H} + \Upsilon_{a} )\right|_{T_{\rm osc}^B}$ condition.\\
\vskip -0.25cm
\noindent Phase-{\bf III} ($T\leq T_n$): The Higgs $vev$, $v_h(T)$ becomes non zero and $v_s$ vanishes at this phase, thereby modifying the ALP mass by the Higgs-portal interaction with a different amount than the Phase-{\bf II}. Again, by the proper choice of parameter $k$, the oscillation frequency of the ALP in this phase can be changed than the previous, hence affecting the final relic abundance.\\
\begin{figure}[htb!]
	 \includegraphics[width=\columnwidth]{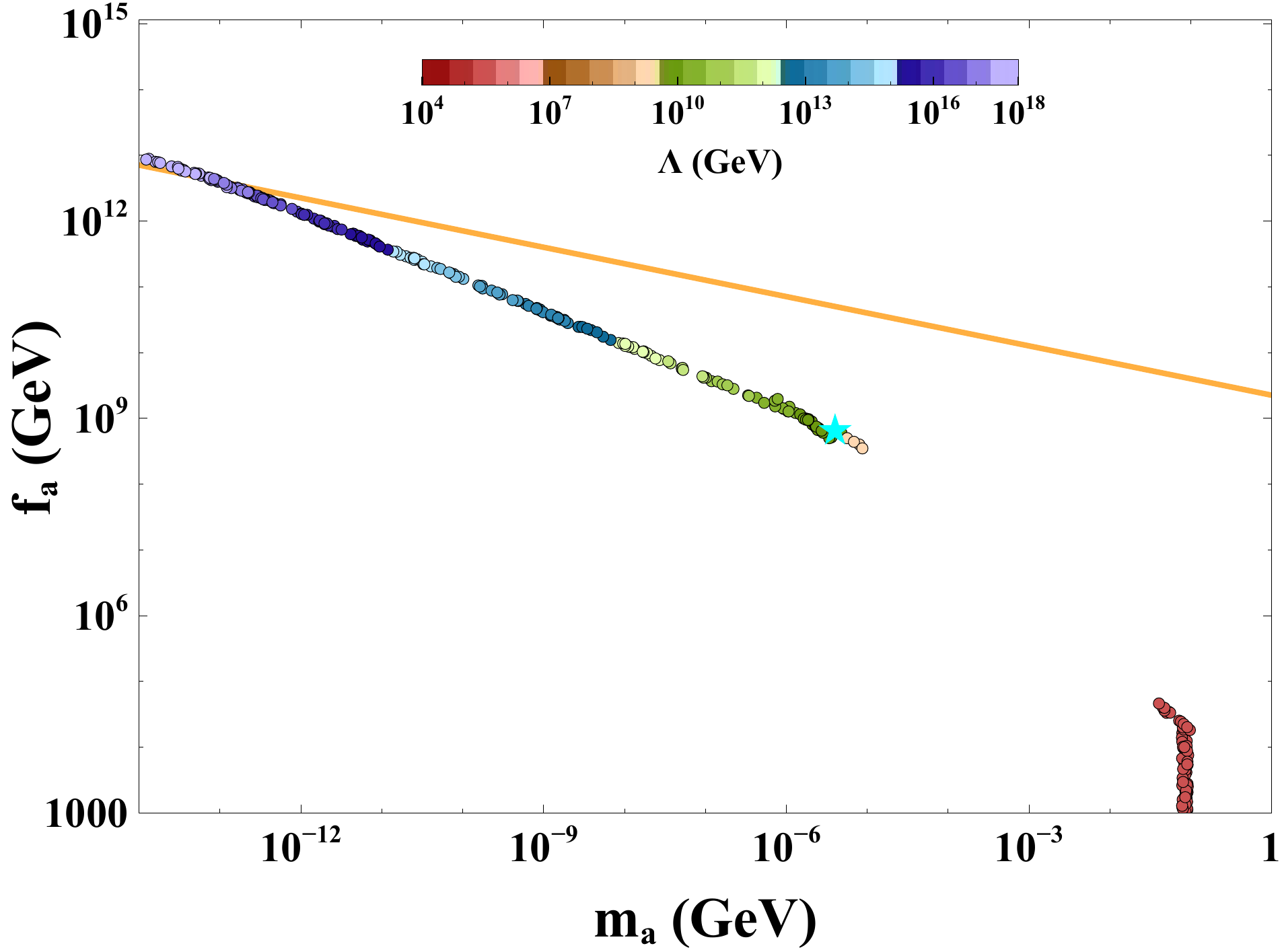}
	\caption{ALP dark matter parameter space in $m_a-f_a$ plane satisfying the correct relic abundance for Case-B. The cyan coloured star indicates the benchmark BP-II in Table~\ref{tab:Two BP-baryonasymmetry-DM relic}, satisfying both the observed BAU and the correct relic abundance.
		The colorbar signifies the variation of the cut-off scale, $\Lambda$.  }
	\label{fig:ALP-ps-B}
\end{figure}

Similar to Case-A, the non-standard evolution of ALP in this case significantly impacts the parameter space in $m_a-f_a$ plane as shown in the Fig. \ref{fig:ALP-ps-B}. However, unlike in case-A, the ALP parameter space, satisfying correct dark matter relic density, does not get enlarged compared to the parameter space in the standard case (orange line). This is because, in this case, ALP cannot begin oscillating at $T>T_s$ due to its extremely small initial mass ($m_{a0}$). Rather, it starts to oscillate mainly at $T\lesssim T_s$ with a higher frequency due to a sudden mass gain via $S$-{\it vev}, $v_s(T)$. Thus, ALP starts to oscillate with a mass, $m_a\simeq v_s(T)^2/\Lambda$. Although $v_s$ gradually vanishes at $T=T_n$, the ALP mass still remain significantly large due to the other dimension-6 operator, assisted by the SM Higgs boson. One can notice that towards the left edge of the figure, as $\Lambda$ approaches $M_{\rm Pl}$, the effect of the dimension-6 term $( v_s(T)^2/\Lambda)$ becomes diluted and ALP starts to oscillate at a temperature connecting with its initial mass $m_{a0}$ upon satisfying the condition $m_{a0}=3\mathcal{H}(T)$. Consequently, the parameter space smoothly merges with the standard case contour in this region. Towards the right edge of the figure, one can see a sharp drop like in the previous Case-A and can be explained through the same line of reasoning as well. Also, we can see a gap in the parameter space in this case because to satisfy the relic we need $\Lambda< 2 \pi f_a$ in this gap region and that's why it is excluded for our scenario.

\section{Baryogenesis}

Having discussed the evolution of ALP as dark matter and its phenomenology in presence of FOEWPT, we now move on to investigate the possibility of generating the baryon asymmetry of the Universe via spontaneous EWBG. Note that the damped coherent oscillation of the ALP, necessary for emerging as dark matter of the Universe, allows us to implement the spontaneous baryogensis at the electroweak scale to solve the matter-antimatter asymmetry problem once we consider the ALP to couple with electroweak anomaly via Eq. \ref{EW anomaly-ALP}. 
Due to the fact that baryon $(B)$ and lepton $(L)$ currents are anomalous in the SM, the evolution of baryon number density can be written as~\cite{Im:2021xoy}
\begin{equation}
	\frac{dn_B}{dt} = N_g \left<  \frac{1}{16 \pi^2} \frac{a}{f_a} {\rm Tr} W^{\mu\nu}\tilde{W}_{\mu\nu} \right> , \label{eq: baryon number evolution}
\end{equation}
where $N_g=3$ is the number of generations. The right hand side (RHS) of Eq.~\ref{eq: baryon number evolution} represents the thermal average of the Chern-Simons (CS) number density which can be estimated as~\cite{Moore:1996qs,Berghaus:2020ekh}
\begin{equation}
	\left< \frac{1}{16 \pi^2} \frac{a}{f_a} {\rm Tr} W^{\mu\nu}\tilde{W}_{\mu\nu} \right> = \frac{\Gamma_{\rm sp}}{2 T}  \frac{d}{dt} \left( \frac{a}{f_a} \right) - \frac{13}{4} \frac{\Gamma_{\rm sp}}{T^3} n_B . \label{eq: CS number evolution}
\end{equation}
Here, $\Gamma_{\rm sp}\simeq 18 \alpha_{\rm W}^5 T^4$ represents the electroweak sphaleron rate per unit volume~\cite{DOnofrio:2014rug}, violating the baryon number. The first term in the RHS of Eq.~\ref{eq: CS number evolution} proportional to $da/dt$ (related to the ALP velocity) serves as the effective chemical potential of baryon number which sources the baryon asymmetry generation while the second term is a wash-out factor.
\begin{figure}[htb!]
	\hspace{-0.6cm}	\includegraphics[width=\columnwidth]{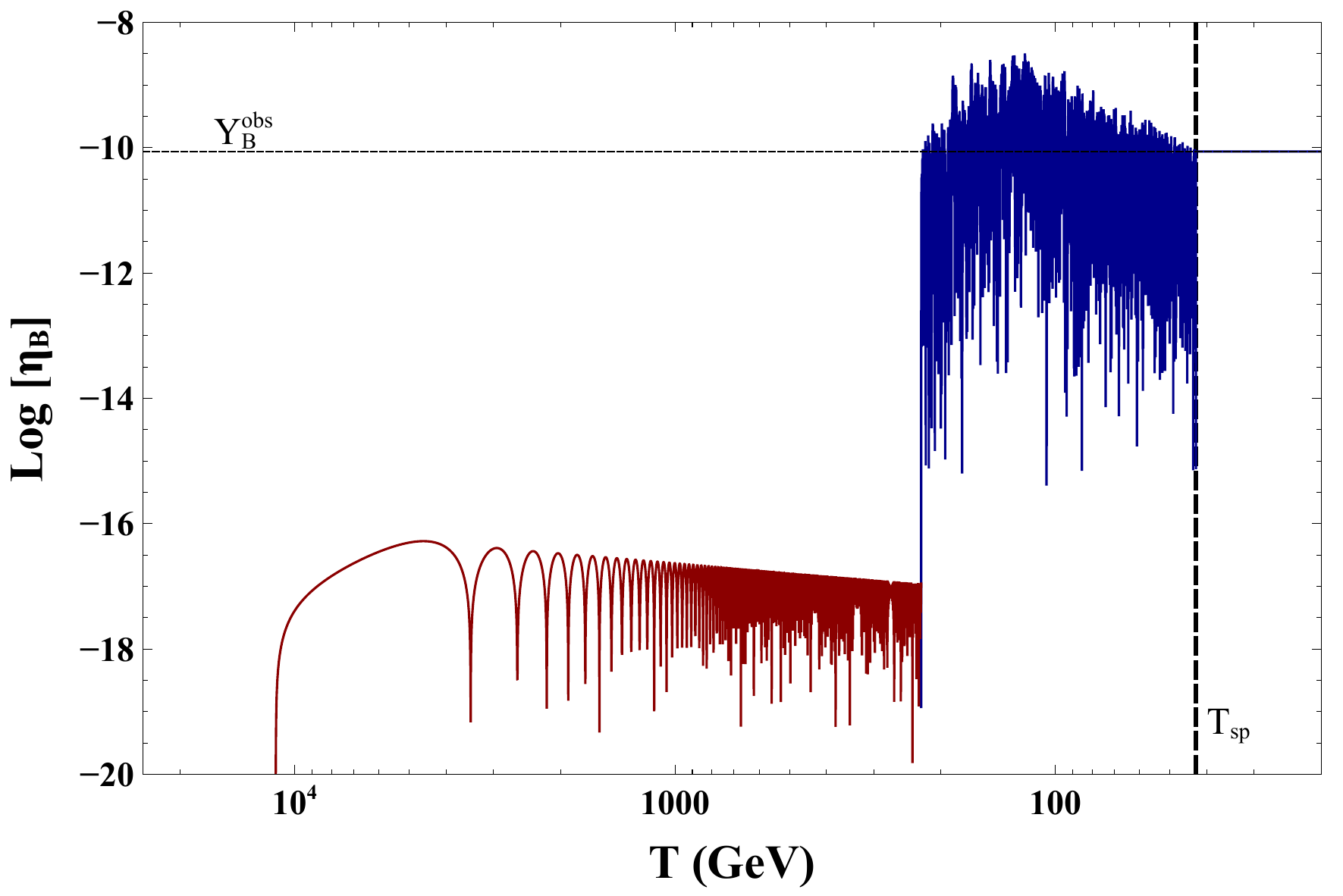}
	\caption{The baryon asymmetry generation driven by ALP oscillation for Case-A.}
	\label{fig:ALP-etaB}
\end{figure}
In this scenario, the baryon asymmetry is generated through the Eq.~\ref{eq: baryon number evolution}, can be expressed in the form as
\begin{equation}
	\frac{dn_B}{dt} = N_g \frac{\Gamma_{\rm sp}}{2 T}  \frac{d}{dt} \left( \frac{a}{f_a} \right) - \frac{13 N_g}{4} \frac{\Gamma_{\rm sp}}{T^3} n_B .   \label{eq: Final baryon asymmetry evolution}
\end{equation}
In our scenario, like as the dark matter case, we have discussed the baryon asymmetry generation in both the cases of the ALP oscillation begins above $T_s$ and below $T_s$. 

\noindent{\underline {Case\text{-}A}:} In this case, the ALP begins oscillating at a temperature higher than $T_s$ (as determined by Eq.~\ref{eq:caseA-Tosc}). Consequently, it starts generating baryon asymmetry according to Eq.~\ref{eq: Final baryon asymmetry evolution}, since $\dot{a} \neq 0$, as illustrated by the red region in Fig.~\ref{fig:ALP-etaB}. However, the resulting asymmetry is insufficient to account for the observed value. Thus, in the standard scenario with this ALP mass, the matter--antimatter asymmetry of the Universe cannot be explained. Fortunately, the enhancement of the ALP mass in the temperature window $T_s \geq T > T_n$, induced by the nonzero vacuum expectation value of the $S$ field during the EWPT, alters the oscillation frequency of the ALP in this regime and boosts baryon asymmetry production. This enhancement, shown by the blue region in Fig.~\ref{fig:ALP-etaB}, allows the mechanism to reproduce the correct observed asymmetry.

\noindent{\underline{Case\text{-}B}:} In this scenario, the ALP would begin oscillating at a temperature below the nucleation temperature $T_n$, based solely on its bare mass $m_{a0}$. Consequently, in the standard setup, no baryon asymmetry can be produced, since sphaleron transitions, the baryon number-violating processes, are already decoupled for $T < T_n$. However, in our case, the enhancement of the ALP mass in the window $T_s \geq T > T_n$ triggers ALP oscillations within this temperature range, thereby enabling the generation of the observed baryon asymmetry as the sphaleron is in equilibrium at this temperature window.

We present the two benchmark points for ALP corresponding to Case-A and Case-B, satisfying the correct observed baryon asymmetry of the Universe $Y_{B}^{\rm obs}\simeq 8.7 \times 10^{-11}$~\cite{Planck:2018vyg} as well as satisfying the correct dark matter relic abundance in the Table~\ref{tab:Two BP-baryonasymmetry-DM relic}.
\renewcommand{\arraystretch}{1.4}
 \begin{table}[!htb]
	\begin{tabular}{| c | c | c | c | c |}
		\hline
		& $m_{a0} $ (GeV)  & $m_a$ (GeV) & $f_a$ (GeV) & $\Lambda$ (GeV)  \\
		\hline
		BP-I & $1.43\times 10^{-10}$ &  $1.065 \times 10^{-8}$  &  $6.5 \times 10^{10}$  &  $1.8 \times 10^{12}$    \\
		\hline
		BP-II & $ 10^{-16}$ &  $4.26 \times 10^{-6}$  &  $6.52 \times 10^{8}$  &  $4.5 \times 10^{9}$ \\
		\hline
	\end{tabular}
	\caption{Benchmark points satisfying correct baryon asymmetry of the Universe as well as correct dark matter relic abundnce.}
	\label{tab:Two BP-baryonasymmetry-DM relic}
\end{table}\\
The two benchmark points, BP-I and BP-II, are shown as cyan stars in the ALP dark matter compatible parameter space plots (in $m_{a}-f_a$ plane) in Fig.~\ref{fig:ALP-ps-A} and Fig.~\ref{fig:ALP-ps-B}, respectively.

\section{Gravitational Wave Spectrum}
\begin{figure}[htb!]
	\hspace{-0.6cm}	\includegraphics[width=\columnwidth]{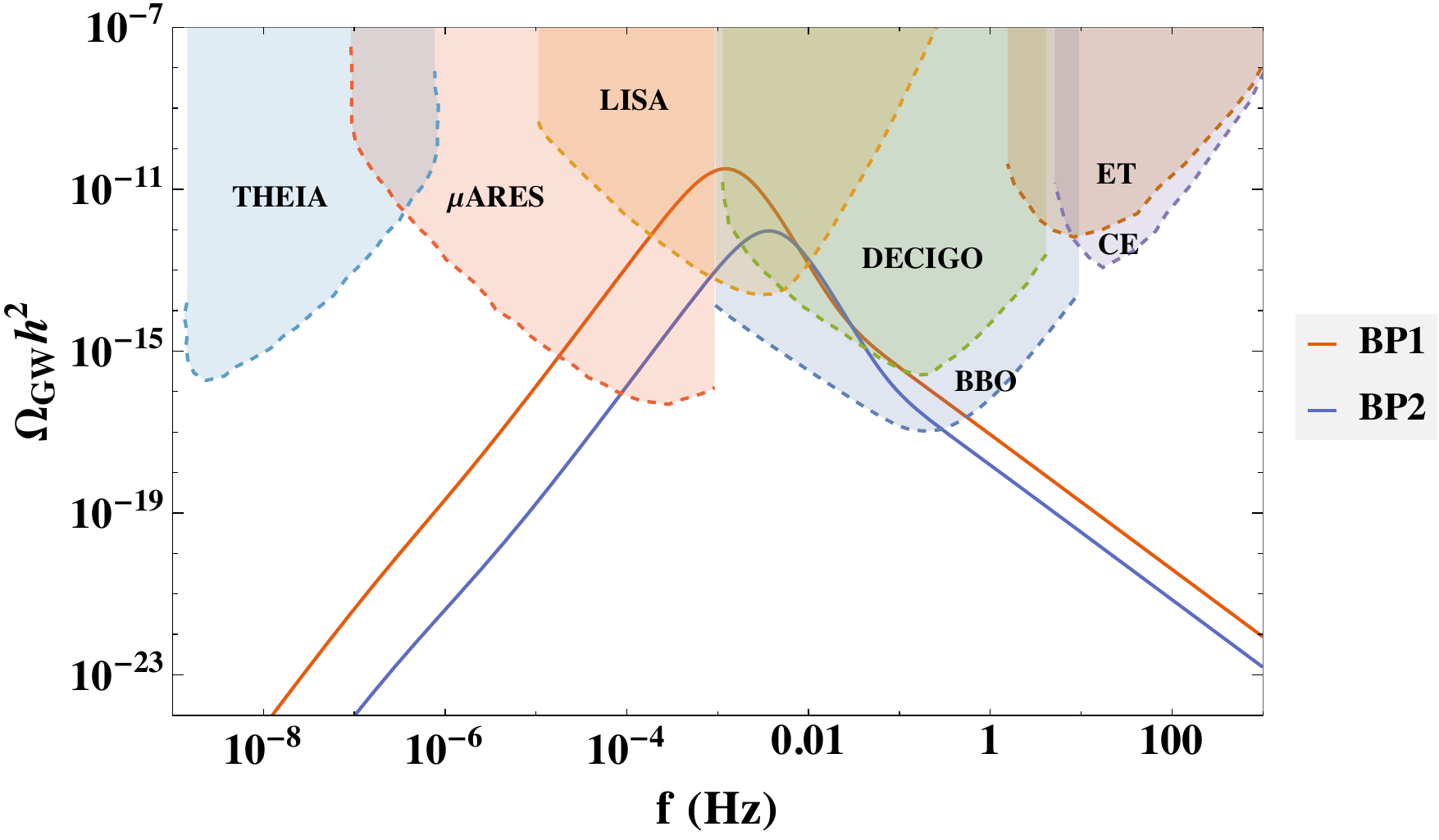}
	\caption{Gravitational Wave spectrum for FOEWPT}
	\label{fig:GW-spectrum}
\end{figure}

In our scenario, the first-order EWPT plays a central role by enhancing the ALP mass prior to the nucleation temperature $T_n$. This enhancement allows the ALP to acquire sufficient velocity, enabling the generation of the baryon asymmetry through its coupling to the EW anomaly. At the same time, the FOEWPT induces nontrivial temperature-dependent dynamics for the ALP, which in turn broadens the viable parameter space for ALP dark matter. Moreover, the occurrence of a FOEWPT naturally leads to the production of a stochastic gravitational wave background, providing a complementary observational probe. Thus, baryogenesis, dark matter phenomenology, and gravitational wave signatures are interconnected within this framework. The resulting gravitational wave spectrum (see Appendix~\ref{ap:GW-Spectrum} for formulae) is shown in Fig.~\ref{fig:GW-spectrum} for the two benchmark points listed in Table~\ref{tab:Two Benchmark}.
As illustrated, the predicted signal lies within the prospective sensitivity ranges of future gravitational-wave observatories such as LISA \cite{LISA:2017pwj}, BBO \cite{Yagi:2011wg,Crowder:2005nr,Corbin:2005ny,Harry:2006fi}, DECIGO \cite{Yagi:2011wg,Kawamura:2006up}, $\mu$ARES \cite{Sesana:2019vho}, CE \cite{LIGOScientific:2016wof,Reitze:2019iox}, ET \cite{Punturo:2010zz,Hild:2010id,Sathyaprakash:2012jk,ET:2019dnz}, and THEIA \cite{Garcia-Bellido:2021zgu}.  
\section{Conclusion}

A successful explanation of the existing matter-antimatter asymmetry requires new dynamics beyond the Standard Model, and electroweak baryogenesis remains one of the most appealing mechanisms. However, realizing EWBG demands a first-order electroweak phase transition (EWPT) together with a dynamical source that biases baryon (or lepton) number during the transition. An axionlike particle provides an attractive option in this context, as its coupling to the electroweak anomaly can naturally supply such a source. Motivated by this, we examine whether a first-order EWPT can trigger a nontrivial evolution of an ALP that not only drives EWBG but also accounts for a dark matter candidate over a wider parameter space than in the usual misalignment picture.

In this work, we focus on a scenario in which the FOEWPT is triggered by a SM singlet scalar field $S$. Interestingly, the temporary non-zero vacuum expectation value of such scalar leads to a transient enhancement of ALP mass in our scenario via non-renormalizable explicit $U(1)$-breaking interactions. This short-lived deformation of the ALP potential modifies its oscillation frequency, enlarging the viable dark matter parameter space. Although we introduced this idea previously with QCD axion, allowing it to reproduce the correct relic abundance for a wide range of $f_a$, the strict relation between the mass and decay constant for the QCD axion (unlike ALPs) prevents it from participating in spontaneous EWBG within this framework. In contrast, for an ALP, the sudden mass increase generated by the non-renormalizable operators accelerates the field and produces a short but substantial burst of ALP velocity, with additional freedom provided by its independent decay constant and cutoff scale. Via the ALP coupling to the electroweak anomaly, this time-dependent motion acts as the required source for electroweak baryogenesis, enabling dynamical generation of the baryon asymmetry during the transition while the electroweak sphaleron is in equilibrium. Altogether, the inclusion of higher-dimensional $U(1)$-breaking terms not only broadens the ALP dark matter parameter space into experimentally accessible regions but also links ALP cosmology to electroweak baryogenesis, while predicting a complementary stochastic gravitational wave background from the underlying first-order EWPT.

\appendix
\onecolumngrid

\section{Construction of the Finite-Temperature Effective Potential} \label{ap:effective-potential}
In this appendix, we briefly outline the steps involved in building the finite-temperature effective potential. As mentioned previously, we introduce a real singlet scalar $S$, odd under the $Z_2$ symmetry, to induce a first-order electroweak phase transition through its coupling to the Higgs field. The corresponding tree-level potential, written in terms of the classical background fields $h$ and $S$, is given in Eq. \ref{tree-level potential} as

\begin{align}
	V_0(h,S) =\, &  \frac{1}{2} \mu_h^2  h^2 + \frac{\lambda_h}{4} h^4 +  \frac{1}{2}  \mu_s^2  S^2 + \frac{\lambda_s}{4} S^4  +  \frac{\lambda_{hs}}{4} h^2 S^2 .  
\end{align}

\noindent To analyze the phase transition, we have to include the zero-temperature one-loop correction, the one-loop thermal contribution, and the higher loop ring (daisy) resummation effects~\cite{Arnold:1992rz,Carrington:1991hz} on top of the tree-level potential. The full effective potential can thus be written as
\begin{align}
	V_{\rm eff}(h,S,T) = V_{0}(h,S) + V_{\rm CW}(h,S) & + V_{\rm T}(h,S,T) +  V_{\rm daisy}(h,S,T) , \label{ap-effective-potential}
\end{align}
where, the one-loop thermal correction is expressed as~\cite{Quiros:1999jp}
\begin{equation}
	V_{T}(h,S,T)=\frac{T^4}{2\pi^2} \left[\sum_{i=h,S,G, W,Z}n_i J_B\left(\frac{m_i^2(h,S)}{T^2}\right)-\sum_{i=t} n_i J_F\left(\frac{m_i^2(h)}{T^2}\right) \right], \label{thermal-potential-1}
\end{equation}
with $J_{B,F}$ being the thermal functions, defined as
\begin{equation}
	J_{B,F}\left(\frac{m_i^2}{T^2}\right)=\int_0^{\infty} dx x^2 \log\left[1\mp {\rm exp}\left(-\sqrt{\frac{x^2+m_i^2}{T^2}}\right)\right]. \label{thermal-integral-1}
\end{equation}
Here $\mp$ signs are for bosons and fermions, respectively. The factors 
\begin{align}
	n_s=1, \,\, n_h=1, \,\,  n_G=3, \,\, n_W=6, \,\,  n_Z=3,  \,\,  n_t=12 ,
\end{align} 
indicate the degrees of freedom correspond to the $S$ field, the Higgs, the Goldstone bosons (in the SM Higgs sector), $W$ boson, $Z$ boson and the top quark, respectively. The field dependent masses $m_i$ appearing in the effective potential are given by
\begin{align}
	m_h^2(h,S) & = \mu_h^2 +3 \lambda_h h^2 + \frac{\lambda_{hs}}{2} S^2  \nonumber \\ 
	m_G^2(h,S) & =  \mu_h^2 + \lambda_h h^2 + \frac{\lambda_{hs}}{2} S^2   \nonumber \\
	m_s^2(h,S) & = \mu_s^2 + 3 \lambda_s S^2 + 	\frac{\lambda_{hs}}{2} h^2  \nonumber \\
	m_{W}^2(h) & =  \frac{g_{\rm w}^2}{4} h^2, \quad 
	m_{Z}^2(h) = \frac{g_{\rm w}^2 + g_{Y}^2}{4} h^2, \quad  
	m_t^2(h)  = \frac{y_t^2}{2} h^2,  \nonumber
\end{align}
where $g_{\rm w}$ and $g_Y$ are the electroweak gauge couplings, and $y_t$ is the top quark yukawa coupling. \\
Here, the one-loop Coleman–Weinberg correction~\cite{Coleman:1973jx} is neglected, since it produces only a small shift in the vacuum structure for the parameter space of interest. The effective potential is therefore constructed using the high-temperature expansion of the thermal functions. In this limit, the bosonic and fermionic thermal integrals take the form
\begin{equation}
	J_B(m^2/T^2)  = - \frac{\pi^4}{45} +  \frac{\pi^2}{12}  \frac{m^2}{T^2} -  \frac{\pi}{6}  \left(\frac{m^2}{T^2}\right)^{3/2} + ... \hspace{0.1 cm}, \quad J_F(m^2/T^2) = -  \frac{7\pi^4}{360} -  \frac{\pi^2}{24}  \frac{m^2}{T^2} + ...  \hspace{0.1 cm} .
\end{equation}
Retaining terms up to order $T^2$ and therefore omitting daisy resummation, the effective potential simplifies to
\begin{align}
	V_{\rm eff}(h,S,T) =  \frac{1}{2}( \mu_h^2 + c_h T^2) h^2 + \frac{\lambda_h}{4} h^4 + \frac{1}{2}( \mu_s^2 + c_s T^2) S^2 + \frac{\lambda_s}{4} S^4 +  \frac{\lambda_{hs}}{4} h^2 S^2 , \label{ap-effective-high-T-potential}
\end{align}
where, the coefficients $c_h$ and $c_s$ capture the leading thermal corrections to the Higgs and singlet masses arising from loops of bosons and fermions that couple to the respective fields. This potential reflects the characteristic vacuum structure of the model: at high temperatures the minimum typically lies along the singlet direction, while at lower temperatures the electroweak vacuum with $\langle h\rangle\neq 0$ becomes a global minimum. Because these two minima lie in different field directions, the mixed quartic interaction $\frac{\lambda_{hs}}{4} h^2 S^2$ prevents a smooth interpolation between them, generating a tree-level barrier in field space, responsible for inducing a FOPT. We therefore utilize this simplified high-temperature potential throughout the main text to analyze the thermal evolution of the vacuum structure and the resulting electroweak phase transition.


\section{Euclidean action and Nucleation temperature determination}\label{ap:Action-calculation}

The decay of the false vacuum proceeds through bubble nucleation, with a decay rate~\cite{Coleman:1977py,Linde:1980tt,Linde:1981zj}
\beeq
\Gamma_d(T) \simeq T^4 \left( \frac{S_3(T)}{2\pi T} \right)^{3/2} {\rm exp}(-S_3(T)/T)
\eeq
To compute this quantity, we evaluate the three-dimensional Euclidean action $S_3$ associated to an $O(3)$-symmetric bubble configuration~\cite{Linde:1981zj},
\begin{align}
	S_3(T) = 4\pi \int dr_E r_E^2  \left(   \frac{1}{2}  \left( \frac{dh}{d r_E}\right)^2 +  \frac{1}{2}  \left( \frac{dS}{d r_E}\right)^2  + V_{\rm eff}(h,S,T)  \right) ,
\end{align}
where $r_E$ is the radial Euclidean coordinate, measured from the center of the bubble. This field configuration that interpolates between the true and false vacua is also known as the bounce solution~\cite{Coleman:1977py}, which satisfies
\begin{equation}
	\frac{d^2\phi_i}{d r_E^2} +  \frac{2}{r_E}  \frac{d\phi_i}{d r_E} -  \frac{d V_{\rm eff}}{d \phi_i}=0, \label{ap:bounce-equation}
\end{equation}
for $\phi_i =\{h,S\}$, together with the boundary conditions
\begin{equation}
	\phi_i(r_E\rightarrow \infty)=(0,v_s(T)) \quad {\rm and}  \quad \left. \frac{d\phi_i}{d r_E}\right|_{r_E=0}=0,
\end{equation}
where $(0,v_s(T))$ denotes the false minima.

We solve this Eq.~\ref{ap:bounce-equation} numerically using the Mathematica based Package {\bf FindBounce}~\cite{Guada:2020xnz}, and evaluate the resulting action $S_3(T)$. The nucleation temperature $T_n$ is defined as the temperature at which bubble formation becomes efficient, which is well approximated by the condition
\begin{align}
	\frac{S_3(T_n)}{T_n} \simeq 140.  \label{action-140}
\end{align}
This criterion follows from equating the false-vacuum decay rate $\Gamma_d(T)$ to the Hubble expansion rate $\mathcal{H}(T)$ during radiation domination. 

\section{Calculation of the Gravitational Wave Spectrum} \label{ap:GW-Spectrum}
Cosmological first-order phase transitions generate stochastic gravitational waves (GWs) through three primary mechanisms: bubble collisions, acoustic (sound) waves in the plasma, and magnetohydrodynamic (MHD) turbulence. The relative importance of these sources depends on the microphysics of the transition, in particular the wall velocity $v_w$ of the expanding bubbles.

In this work we focus on the scenario in which the bubble wall velocity is subsonic, $v_w<c_s$ in the plasma, with $c_s=1/\sqrt{3}$ denoting the sound speed. For such deflagration-type transitions, the bubble wall does not accelerate indefinitely; instead, most of the released vacuum energy is transferred into bulk fluid motion rather. As a result, the dominant GW sources are sound waves and MHD turbulence. The total spectrum can therefore be written as~\cite{Caprini:2015zlo}
\begin{equation}
	\Omega_{\rm GW}h^2 \simeq \Omega_{\rm sw} h^2 + \Omega_{\rm turb} h^2.
\end{equation}
The contributions from sound waves and MHD turbulence take the form~\cite{Athron:2023xlk,Guo:2020grp,Caprini:2019egz,Hindmarsh:2020hop}
\begin{align}
	\Omega_{\rm sw}(f) h^2 & =\frac{1.23 \times 10^{-5}}{g_*^{1/3}} \frac{H_*}{\beta} \left( \frac{k_{\rm sw} \alpha_*}{1+\alpha_*} \right)^2 v_w S_{\rm sw}(f) \Upsilon , \\
	\Omega_{\rm turb}(f) h^2 & =\frac{1.55 \times 10^{-3}}{g_*^{1/3}} \frac{H_*}{\beta} \left( \frac{k_{\rm turb} \alpha_*}{1+\alpha_*} \right)^{\frac{3}{2}} v_w S_{\rm turb}(f).
\end{align}
Here, the factor $\Upsilon = 1 - \frac{1}{\sqrt{1 + 2 \tau_{\rm sw}  H_*}}$ denotes a suppression factor~\cite{Guo:2020grp}, where $\tau_{\rm sw} \sim R_*/\bar{U_f} $ is the sound-wave duration, with $R_* = (8\pi)^{1/3} v_w \beta^{-1}$ is the mean bubble separation and $\bar{U_f}= \sqrt{\frac{3 k_{\rm sw} \alpha_*}{4(1 + \alpha_*)}} $ is the rms fluid velocity.\\
The quantity $\beta/H_*$ describes how fast the transition completes relative to the Hubble expansion rate. Under the assumption that GWs are produced near the nucleation temperature $T_n$, it can be evaluated as~\cite{Grojean:2006bp}
\begin{equation}
	\frac{\beta}{H_*} \simeq T_n \left. \frac{d}{dT}\left( \frac{S_3}{T} \right). \right|_{T=T_n}
\end{equation}

\noindent The strength parameter $\alpha_*$, measuring the released vacuum energy normalized to the radiation energy density at the moment of GW production (at $T=T_n$), is given by~\cite{Grojean:2006bp}
\begin{equation}
	\alpha_* = \frac{1}{\rho^*_{\rm rad}} \left[ \Delta V - \frac{T}{4} \frac{\partial \Delta V}{\partial T} \right]_{T=T_n},
\end{equation}
with $\Delta V = V_{\rm eff}(0,v_s(T),T)-V_{\rm eff}(v_h(T),0,T)$ and $\rho^*_{\rm rad} = g_* \pi^2 T_n^4/30$ being the radiation energy density at $T_n$.

\noindent The spectral shapes for sound-wave and turbulence contributions are parameterized as~\cite{Caprini:2015zlo,Vaskonen:2016yiu}
\begin{align}
	S_{\rm sw}(f) = \left( \frac{f}{f_{\rm sw}} \right)^3 \left( \frac{7}{4+3(f/f_{\rm sw})^2}  \right)^{7/2} ,\quad
	S_{\rm turb}(f) = \frac{(f/f_{\rm turb})^3}{(1+ (f/f_{\rm trub}))^{\frac{11}{3}} (1+ 8\pi f/h_*)},
\end{align}
with 
\begin{equation}
	h_*=1.65 \times 10^{-5} \hspace{0.04 cm}{\rm Hz} \left( \frac{T_n}{100 \hspace{0.04 cm} {\rm GeV}} \right)  \left( \frac{g_*}{100 } \right)^{\frac{1}{6}}
\end{equation}
being the redshifting factor. The peak frequencies, $f_{\rm sw}$ and $f_{\rm turb}$, of each contribution can be written as
\begin{align}
	f_{\rm sw}  = \frac{1.9 \times 10^{-5} \hspace{0.04 cm}{\rm Hz}}{v_w} \frac{\beta}{H_*} \left( \frac{T_n}{100 \hspace{0.04 cm} {\rm GeV}} \right)  \left( \frac{g_*}{100 } \right)^{\frac{1}{6}} , \quad
	f_{\rm turb} = 1.42 f_{\rm sw} .
\end{align}

\noindent The efficiency factors $k_{\rm sw}$ and $k_{\rm turb}$ quantify the fraction of released vacuum energy converted into bulk kinetic energy of the fluid and into turbulence, respectively. For subsonic wall velocities, these can be defined as~\cite{Espinosa:2010hh}
\begin{align}
	k_{\rm sw}  = \frac{c_s^{11/5} k_{\rm a}k_{\rm b}}{(c_s^{11/5}-v_w^{11/5})k_{\rm b} + v_w c_s^{6/5} k_{\rm a}} , \quad
	k_{\rm turb}  = \epsilon k_{\rm sw} ,
\end{align}
with $\epsilon$ is typically in the range of $5\%$-$10\%$~\cite{Caprini:2015zlo,Hindmarsh:2015qta}. We take $\epsilon=0.05$ as a conservative choice. $k_{\rm a}$ and $k_{\rm b}$ are defined as~\cite{Espinosa:2010hh}
\begin{align}
	k_{\rm a} = \frac{6.9 v_w^{6/5}\alpha_*}{1.36 - 0.037 \sqrt{\alpha_*} + \alpha_*} ,\quad
	k_{\rm b} = \frac{\alpha_*^{2/5}}{0.017 + (0.997 + \alpha_*)^{2/5}}.
\end{align}
This framework provides the full gravitational-wave spectrum generated by a subsonic first-order electroweak phase transition, and it is the expression used in the main analysis of this work.

\twocolumngrid

\bibliography{ref-ALP.bib}

\end{document}